\documentclass[11pt,a4paper]{article}
\def\bea{\begin{eqnarray}}
\def\eea{\end{eqnarray}}

\def\beq{\begin{equation}}
\def\eeq{\end{equation}}
\def\ba{\beq\new\begin{array}{c}}
\def\ea{\end{array}\eeq}
\def\be{\ba}
\def\ee{\ea}

\makeatletter
\newdimen\normalarrayskip % skip between lines
\newdimen\minarrayskip % minimal skip between lines
\normalarrayskip\baselineskip
\minarrayskip\jot
\newif\ifold \oldtrue \def\new{\oldfalse}

\def\arraymode{\ifold\relax\else\displaystyle\fi} % mode of array entries
\def\eqnumphantom{\phantom{(\theequation)}} % right phantom in eqnarray
\def\@arrayskip{\ifold\baselineskip\z@\lineskip\z@
\else
\baselineskip\minarrayskip\lineskip2\minarrayskip\fi}

\def\@arrayclassz{\ifcase \@lastchclass \@acolampacol \or
\@ampacol \or \or \or \@addamp \or
\@acolampacol \or \@firstampfalse \@acol \fi
\edef\@preamble{\@preamble
\ifcase \@chnum
\hfil$\relax\arraymode\@sharp$\hfil
\or $\relax\arraymode\@sharp$\hfil
\or \hfil$\relax\arraymode\@sharp$\fi}}

\def\@array[#1]#2{\setbox\@arstrutbox=\hbox{\vrule
height\arraystretch \ht\strutbox
depth\arraystretch \dp\strutbox
width\z@}\@mkpream{#2}\edef\@preamble{\halign
\noexpand\@halignto
\bgroup \tabskip\z@ \@arstrut \@preamble \tabskip\z@ \cr}
\let\@startpbox\@@startpbox \let\@endpbox\@@endpbox
\if #1t\vtop \else \if#1b\vbox \else \vcenter \fi\fi
\bgroup \let\par\relax
\let\@sharp##\let\protect\relax
\@arrayskip\@preamble}
\def\eqnarray{\stepcounter{equation}%
\let\@currentlabel=\theequation
\global\@eqnswtrue
\global\@eqcnt\z@
\tabskip\@centering
\let\\=\@eqncr
$$
\halign to \displaywidth\bgroup
\eqnumphantom\@eqnsel\hskip\@centering
$\displaystyle \tabskip\z@ {##}$%
\global\@eqcnt\@ne \hskip 2\arraycolsep

$\displaystyle\arraymode{##}$\hfil
\global\@eqcnt\tw@ \hskip 2\arraycolsep
$\displaystyle\tabskip\z@{##}$\hfil
\tabskip\@centering
&{##}\tabskip\z@\cr}
\begingroup\ifx\undefined\newsymbol \else\def\input#1 {\endgroup}\fi

\begin{document}

\setcounter{footnote}{1}
\def\thefootnote{\fnsymbol{footnote}}
\begin{center}
\hfill ITEP-TH-20/01\\ \hfill hep-th/0104250\\ \vspace{0.3in}
{\Large\bf     Noncommutative field theory}\\
\vspace{0.2cm}
{\Large\bf     in  formalism of first quantization.}

\vspace{0.8cm}
\centerline{{\Large A.Ya.Dymarsky}\footnote{ ITEP and MSU, Moscow, Russia; e-mail: dymarsky@gate.itep.ru}}

\end{center}

\abstract{\footnotesize  }

We present a first-quantized formulation of the quadratic non-commutative field theory
in the background of abelian (gauge) field. Even in this simple case the Hamiltonian of a propagating
 particle depends non-trivially on the momentum (since  external  fields depend on  location of the Landau orbit)
so that one can not  integrate out momentum  to obtain a local theory in the second order formalism.
The cases of scalar and spinning particles are considered. A representation for exact propagators is found
  and the result is applied to description of the
Schwinger-type processes (pair-production in  homogenous external field). \\

\bigskip
\setcounter{footnote}{0}
\renewcommand\thefootnote{\arabic{footnote}}

\section{Introduction}

\ \ \ \ The noncommutative (NC) fields theory have been extensively studied recently.
However, most of consideration were based on the second-quantizes field theory.
Here we are going to study the NC theory within the first quantization
approach.

More concretely  we claim that the
unique difference between ordinary and NC theories in the formalism of first
quantization is that the external fields in the NC case depend on ``shifted" coordinates
$q^{\mu}=x^{\mu}-\frac{1}{2}\theta^{\mu\nu}p_{\nu}$, which are
the counterparts  of  coordinates of the Landau orbit for the charged particle in the  magnetic field. Here $p_{\mu}$ are the
momenta, $x^{\mu}$
are the coordinates of  the particle and $\theta^{\mu\nu}$ is the parameter
of noncommutativity (the constant antisymmetric tensor).

    We consider the theories  quadratic in quantum fields, since   only such theories  have first-quantized formulation.

    This paper is organized as follows. We begin with  verifying our claim in the case of scalar particle in the next section.
 In section 3 we apply our results to
the pair-production process. Section 4 constructs the first-quantized theory
of
spinning particle. At last  we end up  with concluding remarks in section 5.

\section{Scalar particle}
\ \ \ \ In this section we consider correspondence between the first and second
quantized description of the  of scalar particle in the example of the scalar QED with external currents.
The main  statement is that the first-quantized theory with the action (in the first order formalism)
\be
S_1=\int dt(p\dot{x}-(p-A(q))^2-J(q)+m^2), \ \ q^{\mu}=x^{\mu}-\textstyle\frac{1}{2}\theta^{\mu\nu}p_{\nu}
\ee
corresponds to the second quantized theory with the action
\be
S_2=\int d^{D}x \ (D_{\mu}\phi^{+}*D_{\mu}\phi - m^{2} \phi^+ * \phi )+J*\phi^+*\phi \   ,  \
 D_{\mu}=\partial_{\mu}-iA_{\mu}*
.\ee

Here $*$ is the Moyal product with the parameter of noncommutativity $\theta^{\mu\nu}$.
We omit the  1-dimensional metric  along the trajectory from the action (1).
It  means that we already fixed the gauge: $t$ is the natural parameter along the trajectory and $t \in [0,T]$,
 where  T is the proper time (the length of  trajectory).
Then, of the whole path integral over 1-dimensional metric, we are left with
an ordinary integral over the proper time with the measure $dT$ (for details see \cite{Polb}).

    We  prove the correspondence between the theories identifying the exact propagators in
 the first and second quantized theories.
The propagators will be obtained as   perturbative series in both cases.
The exact propagator for the first-quantized theory is given by the
formula
\be
G(x,y)=\int\limits_0^{\infty}dT\int\limits_{x(0)=y}^{x(T)=x} Dx(t)Dp(t) e^{iS}.
\ee
 We  demonstrate by  manifest checking that the perturbative  series in
powers of external fields from propagator (3) with action (1) coincides with
the field series. In other words, we  obtain the Feynman rules for the field
theory (2) from the particle theory (1).
(The perturbative theory in the formalism of first quantization is described in \cite{FH}. The relativistic case and
the correspondence to the field perturbation theory were discussed in \cite{MM}.)

    First we represent the path integral (3) as the limit of ordinary
integrals

 \be
 G(x,y)\equiv \int^{\infty}_{0} d T e^{im^2 T} G(x,y,T)   \   \ ,     \   \        \
 G(x,y,T)\equiv \displaystyle{\lim_{\scriptscriptstyle{N \rightarrow \infty}}} (\prod^{N-1}_{i=1}dx_{i} \
G(x_{i+1},x_{i})_{\frac{T}{N}}) G(x_1,y)_{\frac{T}{N}}.
\ee
 $G(x,y)_{\Delta T}$ is as usual   defined up to the first oder in $\Delta T$ only
\be
  G(x,y)_{\Delta T}\! \equiv \int \!  \frac{d^ D p \ e^{ip(x-y)}}{(2\pi)^D} (e^{-ip^{2}\Delta T}\!+\! i(A_{\mu}( x' )p_{\mu}\!+\! A_{\mu}( y'
)p_{\mu}\!-\!
A_{\mu}( x' )A_{\mu}( y' )\! - \! J( x' ))\Delta T) ,  \\
                          {x'}^{\mu}=x^{\mu}-\textstyle{\frac{1}{2}} \theta^{\mu\nu} p_{\nu} \         ,
  \          y'^{\mu}=y^{\mu}-\textstyle{\frac{1}{2}} \theta^{\mu\nu} p_{\nu}.
\ee
 The points $x'$ and $y'$ are split in such a  way that in the limit
$\theta^{\mu\nu}\rightarrow 0 $, the expression (5) becomes  the ordinary
one known in
the first-quantized scalar QED.

    Note that the parameter of noncommutativity $\theta^{\mu\nu}$ is contained only  in  terms, describing
 the interaction with external fields. Therefore,    the free propagator obtained
from (3) with the action (1)  coincides with the ordinary one.
 This is  why in the NC theory the
internal lines remain ordinary. This fact is in coincidence with the field
theory result.

Now we begin to consider the vertex part of Feynman diagrams. To this end, we
 need the following key formula
 \be
<x| f*  |y>=  \int\frac{d^{D}p}{(2\pi)^D} e^{i(p(x-y))}
f(x^{\mu}+\alpha(y-x)^{\mu}-\frac{1}{2}\theta^{\mu\nu}p_{\nu}),
 \ee
for arbitrary $\alpha \in [0,1]$.
Note that the dependence on $\alpha$ vanishes  in (6) after the integration over
$p$. The crucial point here is that $\theta^{\mu\nu}$ is constant  and (6) has
the very simple form.
Yet another useful  formula, which follows from the previous one, is
 \be
<x| (f*g)*  |y>=  \int\frac{d^{D}p}{(2\pi)^D} e^{i(p(x-y))}
f(x^{\mu}-\frac{1}{2}\theta^{\mu\nu}p_{\nu})g(y^{\mu}-\frac{1}{2}\theta^{\mu\nu}p_{\nu}).
 \ee

    It is evident now that the vertex term (formula (5) without the term $e^{-ip^{2}\Delta T}$
which describes   free propagation) corresponds to the NC theory

\be
\int\frac{d^{D}p}{(2
\pi)^D} e^{i(p(x-y))} i( A_{\mu}(x')p_{\mu}+ A_{\mu}(y')p_{\mu}-
A_{\mu}(x')A(y')_{\mu}-J(x'))\Delta T =\\
=-i<x|  2i(A_{\mu}*)\partial_{\mu}\ + \ i(\partial_{\mu}A_{\mu})* \ + \
(A_{\mu}*A_{\mu})* \  + \  J*|y>.
\ee

It is also obvious now how to construct   for  arbitrary  scalar theory the first-quantized formulation
 from the second-quantized one.

\section{Schwinger type processes}
\ \ \ \ One can not calculate the exact propagator for  arbitrary external
field. However     when the external field is homogenous it is possible.
In this section, we calculate the probability   for the pair production (the imaginary  part of the propagator)  in this the background  in
 the formalism of first quantization and compare our result with the field
theory calculation.

    Let us  consider a particle on the noncommutative plane in the classical external field
 $A_{\mu}(q)=\frac{1}{2}B_{\nu \mu}q^{\nu}$ with homogenous
 field's strength $B_{\mu\nu}=const$  and $J=0$. In this case, one can easy calculate the probability  of pair production
  since  the action  (1)  is
quadratic in all variables.  It is
convenient to change variables from $p,q$ to the
 canonical $\pi,y$
\be
\pi_{\mu}=p_{\nu}K^{\nu}_{\mu } \ \, \ y^{\mu}=x^{\nu}(K^{-1})^{ \ \mu}_{\nu}
 \  , \ \  \  K^{\mu}_{\nu}=\delta^{\mu}_{\nu} +\frac{1}{4}B_{\sigma
\mu}\theta^{\sigma \nu} ,
\ee
where $x^{\mu}=q^{\mu}+\frac{1}{2}\theta^{\mu\nu} p_{\nu}$. In the new
variables,
the action can be rewritten  as
\be
S=\int dt(\pi_{\mu}\dot{y}^{\mu}-(\pi_{\mu}-\tilde{A}_{\mu}(y))^2+m^2) \    , \
\ \ \tilde{A}(y)_{\mu}=A_{\mu}( K y)        =\frac{1}{2}B_{\sigma \mu} K^{\sigma}_{\nu} y^{\nu}
.\ee
It is remarkable  that this action corresponds to an ordinary particle  in
external field with the strength
\be \tilde{F}_{\mu \nu}=\frac{\partial \tilde{A}_{\nu}}{\partial y^{\mu} } -\frac{\partial \tilde{A}_{\mu}}{\partial y^{\nu}
}. \ee
Now one can  calculate the probability of pair creation in the usual way. The result (gauge invariant quantity) will depend only on $\tilde{F}_{\mu \nu}$.
At the same time,  $\tilde{F}_{\mu\nu}$  changes under gauge
transformations
\be
 A_{\mu} \rightarrow g*A_{\mu}*g^{+}-i\partial_{\mu}g*g^{+} , \
\ g^+*g=1.
\ee
The solution of this puzzle  is that the NC strength
\be
F_{\mu \nu}=\frac{\partial A_{\nu}}{\partial q^{\mu} } -\frac{\partial A_{\mu}}{\partial
q^{\nu}}-i[A_{\mu},A_{\nu}]_{*}

\ee
is  still gauge invariant for homogenous fields. Moreover,  (a little surprise) $F_{\mu
\nu}$ and $\tilde{F}_{\mu\nu}$ are equal to each other
\be
F_{\mu\nu}=\tilde{F}_{\mu\nu}=B_{\mu\nu}+\frac{1}{4}\theta^{\sigma
\rho}B_{\sigma\mu}B_{\rho\nu}.
\ee

    Therefore, the probability is given by the standard  formula (see \cite{Sch}) but with noncommutative field
strength $\tilde{F}_{\mu\nu}$. For example, in the 4-dimensional space, when the magnetic field vanishes
 ($\tilde{F}_{i,j}=0, \  \ i,j=\overline{1,3}$), the probability is
\be
w=\frac{e^2 |\overrightarrow{E}|^2}{8 \pi ^3}\sum\limits_{n=1}^{\infty}\frac{(-1)^{n+1}}{n^2} exp(-\frac{\pi n
m^2}{|e\overrightarrow{E}|}),  \  \ \ eE_i=\tilde{F}_{0 i}.
\ee
  This result certainly coincides with the field theory calculation (see \cite{AGB,SJ}).

\section{Spinning particle}
\ \ \ \ Now we are going to construct the  exact
propagator  for the spinning particle in the NC case similar to what we did below.

First, we rederive   the analogue of formula (3) for the spinning particle
 and calculate the first correction in external fields  to the exact propagator in the ordinary (commutative) theory.
 In this section, we work in the Euclidean space for simplicity.
We start from another representation for the propagator of scalar particle (see
\cite{Polb})
 \be
G(p_2,p_1)=\int\limits_{0}^{ \infty } \frac{d T e^{-mT}}{(2 \pi)^{2D}} \int Dx(t) e^{ip_2 x(T)}
\delta(\dot{x}^{2}-1)e^{\int A\dot{x}} e^{-ip_1 x(0)}.
\ee
In further consideration  we mainly follow the approach of spin quantization  developed in
\cite{AlSh,AlFSh,NR} (see also \cite{Pol}). In accordance with their approach, in order to
quantize the spin one needs to add  a term $e^{i S_{spin}}$ to (16). Note that
the delta-function  $\delta(\dot{x}^2-1)$ keeps the vector $\dot x ^{\mu}$ in the
sphere. Thus, in fact, in (16) one sums over the trajectories lying  on the sphere (this fact is important).
The term $S_{spin}$ is the integral  of a special external 1-form  connected with the $SO(D)$ group, along such
a trajectory.
Note that this sphere is the configuration space for the  particle's velocity $n^{\mu}= \dot x ^{\mu}$
(or the phase space for the spin), not for the
particle. In order to distinguish these spaces, we denote the space-time
trajectory via $x(t)$, and ``spin" space trajectory via $\eta(t)$. We also  use the convenient variables
$n^{\mu}=\dot \eta ^{\mu}$, $Dn=D \eta \  \delta(\dot \eta ^2-1)$.  Then the exact propagator for spinning particle has the form
\be
G(p_2,p_1)=\int\limits_{0}^{ \infty } \frac{d T e^{-mT}}{(2 \pi)^{2D}} \int D\eta(t) e^{ip_2 \eta(T)}
\delta(\dot{\eta}^{2}-1)e^{iS_{spin}}e^{\int A_{\mu}\dot{\eta}^{\mu}} e^{-ip_1 \eta(0)}=\\
=\int\limits_{0}^{ \infty }  \frac{d T e^{-mT}}{(2 \pi)^{2D}} \int \limits_{\eta(0)=0} D\eta(t)
 e^{ip_2 \int_{0}^{T} \dot{\eta} dt} \delta(\dot{\eta}^{2}-1) e^{iS_{spin}} e^{\int A_{\mu}\dot{\eta}^{\mu}}\delta(p_2-p_1).
\ee
In this case, the perturbative theory is more complicated  than in the scalar case.
This is why we first demonstrate in detail how the vertex terms emerge from (17). The first  correction in $A_{\mu}$ is
\be
G^{1}(p_2,p_1)=\lim_{ \Delta t \rightarrow 0} \int\limits_{\Gamma} \frac{d T dt e^{-mT}}{(2 \pi)^{2D}} \int\limits
  D\eta(t) e^{ip_2 \eta_{T}}A_{\mu}(  {\eta_{t}} ) (\eta_{t'}-\eta_{t})^{\mu}  \delta( \dot{\eta}^{2}-1)
e^{iS_{spin}}e^{-ip_1 \eta_0} ,\\
t'=t+\Delta t,  \  \  \Gamma:T\in [0,\infty]   , t\in [0,T]
.\ee
In order to transform it to the standard form, we use the following important trick: we multiply (18) by  unity
\be
 1=\int \frac{d\eta'_{t'}dk_2}{(2 \pi)^{D}}  \frac{d\eta'_{t}dk_1}{(2 \pi)^{D}} \ e^{ik_2(\eta_{t'}-\eta'_{t'})} e^{ik_1(\eta_{t}-\eta'_{t})}
,\ee
and rewrite it as
\be
 G^{1}(p_2,p_1)=\lim_{ \Delta t \rightarrow 0}\int\limits_{\Gamma} \frac{d T dt e^{-mT}}{(2
\pi)^{2D}} \frac{dk_1 dk_2 d{\eta'}_t d{\eta'}_{t'}}{(2 \pi)^{2D}}  \times \\
 \int  D\eta e^{i(\int p_2\dot{\eta} +\eta'_{t'}(p_2-k_2) +k_2 \eta_{t'})}A_{\mu}(\eta_{t}')(\eta_{t'}'
-\eta_{t})^{\mu} e^{i(-k_1 {\eta'}_t+ \int  k_1 \dot{\eta}+ \eta_0(k_1-p_1))} \delta( \dot{\eta}^{2}-1)
e^{iS_{spin}}.\ee
We have to remove $e^{ik_2(\eta_{t'}-\eta_{t}')}$ from formula (20), analogously we remove $e^{-ip^2 \Delta T}$ from (8)
before  taking the continuum limit $\Delta t \rightarrow 0$. Only after
this, the first correction acquires  the form corresponding to QED
\be
G^{1}(p_2,p_1)=\int dk_{1} dk_{2} d\eta \ G^{0}(p_2,k_2) e^{ik_2 \eta} A_{\mu}(\eta) \gamma^{\mu} e^{-ik_{1} \eta} G^{0}(k_1,p_1)
.\ee

The important moment is that we remove here the integral  $\int
D\eta\delta(\dot{\eta}^2-1)¥^{iS_{spin}}$ and change all $\dot{\eta}^{\mu}$ for the
gamma-matrices $\gamma^{\mu}$ (see the works \cite{AlSh} and \cite{Pol}).

    Now we demonstrate how to deform  the propagator (17) in the NC case.
 The main difficulty is that in  formula (17) we sum over the trajectories in the special
``spin" space, when the particle propagates  straightforwardly ($p=const$).
However, the terms which depend on $q^{\mu}$ need to be  integrated
over the trajectories in the ``space-time" phase space.
This is why if we want to change the argument of external fields in (17), similarly to
(1), we have to add in (17)  the sum over trajectories in such a space. We
can do this in the following way: represent the path integral as the continuum limit of ordinary
integrals similarly we did with (3). After that,  multiply (17) by  unity
\be
 1=\int \frac{dx dp}{(2 \pi)^{D}} e^{ip(\eta-x)}
\ee
for all $\eta(t)$. It is evident   that the dependencies  of external fields on $\eta$ or on $x$ are equivalent.
   Then, we integrate out
the delta-function and pass from  integrating  over $\eta$  to  integrating
over $n$.
(Note that in a similar way    one can
obtain (3) with the commutative ($\theta^{\mu\nu} =0$) action (1)  from (16), by adding new degrees of freedom.)
After that, we change the arguments of all external fields  from $x^{\mu}$ to $q^{\mu}$.
Since we work with propagator in the momentum representation, the boundary
conditions allow us to change the integration over variables $x^{\mu}$ to the
integration over variables $q^{\mu}$. Finally, one obtains
\be

G=\int\limits_{0}^{\infty} d T e^{-mT} \int Dq(t) Dp(t) Dn(t)
e^{i\int_{0}^{T}(p dq +\frac{1}{2}p \theta d p) } e^{i\int_{0}^{T}p n dt}
e^{\int_{0}^{T}A_{\mu}(q)n^{\mu}dt} e^{iS_{spin}[n(t)]} . \ee

 The perturbation series for this formula corresponds to the NCQED. For example,
 the first correction to the propagator has the form
\be
G^{1}(p_2,p_1)=\int dk_{1} dk_{2} dx \ G^{0}(p_{2},k_2) e^{ik_{2} q}  A_{\mu}(q) \gamma^{\mu}
 e^{i(-k_{1} q+ k_1 \theta (k_2-k_1))} G^{0}(k_1,p_{1})
=\\
=\int dk_{1} dk_{2} d q \ G^{0}(p_{2,k_2}) e^{ik_{2} q}* A_{\mu}(q) \gamma^{\mu} * e^{-ik_{1} q}
G^{0}(k_{1},p_1).
\ee
It is obvious that we have to add $e^{\int_0^T J(q)dt}$ to (23) in order to
obtain the first-quantized formulation of NCQED with external current
\be
S_2=\int d^D x (\bar{\psi}*(D_{\mu}\gamma^{\mu}+J)*\psi).
\ee

\section{Concluding remarks}
We demonstrated that quadratic noncommutative field theory can be described
in terms of particles  similarly to the ordinary case.  However, in this case the particle action depends on the momentum nontrivially
and the theory is non-local.  We also
constructed the exact propagators in the case of scalar and spinning particle in the background of classical
abelian gauge field and current.  The result is applied to the pair-production processes.
\\

Author is grateful to A.Gorsky and K.Selivanov for initiating this work and
discussions, to  A.Alexandrov, D.Melnikov and A.Solovyov for  useful
advices and especially to A.Mironov  and A.Morozov for careful reading the manuscript and support.
\\

    This work was partly supported by the Russian President's grant 00-15-99296,  INTAS grant 01-334   and RFBR grant 01-02-17682-a.

\end{document}